\documentclass[floatfix,twocolumn,showpacs,amsmath,amssymb]{revtex4}
\usepackage{graphicx}
\usepackage{dcolumn}
\usepackage{bm}
\usepackage{mathrsfs,psfrag}
\newcommand{\eq}[1]{Eq.~(\ref{#1})}
\begin{document}

\title{Separation and identification of dominant mechanisms in double 
photoionization}
\author{Tobias Schneider}
\email{tosch@mpipks-dresden.mpg.de}
\author{Peter Leszek Chocian}
\email{pete@spymac.com}
\author{Jan-Michael Rost}
\email{rost@mpipks-dresden.mpg.de}
\affiliation{
Max Planck Institute for the Physics of Complex Systems, \\
N\"othnitzer Stra{\ss}e~38, 01187 Dresden, Germany
}

\date{\today}

\begin{abstract}
Double photoionization by a single photon is often discussed in
terms of two contributing mechanisms, {\it knock-out} (two-step-one)
and {\it shake-off} with the latter being a pure quantum effect.  It is
shown that a quasi-classical description of knock-out and a simple
quantum calculation of shake-off provides a clear separation of the
mechanisms and facilitates their calculation considerably.
The relevance of each mechanism at different photon energies is
quantified
for helium.  Photoionization ratios, integral and singly differential
cross sections obtained by us are in excellent agreement with benchmark
experimental data and recent theoretical results.
\end{abstract}


\maketitle

Our understanding of dynamical processes often rests on isolating
approximate mechanisms which leave characteristic traces in the
measured or computed observables.  A prime example is double
photoionization.  After the initial absorption of the photon by the
primary electron the subsequent redistribution of the energy among the
electrons is often discussed in terms of two mechanisms
\cite{doerner}, knock-out (KO) (sometimes called `two-step-one'
\cite{ko})
and shake-off (SO) \cite{so,dalgarno,aberg}.  The first mechanism
describes the correlated dynamics of the two electrons as they leave
the nucleus where the primary electron has knocked out the
secondary electron in an $(e,2e)$-like process. The second mechanism 
accounts for the fact that absorption of the photon may lead to a sudden
removal of the primary electron. This causes a change in the atomic
field so that the secondary
electron relaxes with a certain probability to an unbound state of the
remaining ${\rm He}^+$
ion, {\it i.e.}, the secondary electron is shaken off.

Apart
from general properties, {\it e.g.}, the prevalence of shake-off at
high photon energies, it is difficult to separate the processes.
However, they are distinct
with respect to their quantum nature: shake-off is a purely quantum
mechanical phenomenon while knock-out dynamics occurs classically as
well as quantum mechanically.  This opens up the possibility to
separate shake-off and knock-out by calculating the latter
(quasi-)classically, provided the quasi-classical approximation to
knock-out is good.  Clearly, the quasi-classical KO mechanism does
not contain any part of SO (which is purely quantum).

The two phases of double photoionization, initial absorption and 
redistribution of the energy, can
be expressed by the relation
\begin{equation}
\label{eq1}
\sigma^{++}_{\rm X} = \sigma_{\rm abs} P^{++}_{\rm X}
\end{equation}
where ${\rm X}$ stands for either shake-off or knock-out.  In the
following, we evaluate $P^{++}_{\rm X}$ for full fragmentation of the
ground state and use the experimental data of Samson {\it et al.}
\cite{sam94} for $\sigma_{\rm abs}$.  We obtain the classical double
escape probability for KO with a classical-trajectory Monte-Carlo
(CTMC) phase space method.   CTMC has been frequently used
for particle impact induced fragmentation \cite{ct1,ct2,ct3,ct4} with
implementations differing typically in the way the phase space
distribution, $\rho(\Gamma)$, of the initial state is constructed.
Details of our approach will be published elsewhere, here
we summarize the important steps only.

   Within our phase space
approach the double escape probability $P^{++}_{\text{KO}}$ is
formally given by
\begin{equation}
P^{++}_{\text{KO}}=\lim_{t\rightarrow\infty}\int{\rm d}\Gamma_{\mathcal
{P}^{++}}\,
\exp((t-t_{{\rm abs}})\mathscr{L}_{\text{cl}})\rho(\Gamma),
\label{intphas}
\end{equation}
with the classical Liouvillian
$\mathscr{L}_{\text{cl}}$
for the full three-body Coulomb system propagated from the time
$t_{\mathrm{abs}}$ of photoabsorption.  The projector $\mathcal
P^{++}$ indicates that we have to integrate only over those parts of
phase space that lead to double escape (the asymptotic final energies
of the two electrons, $\varepsilon$ and $E-\varepsilon$, are
positive).  The integral in Eq.~(\ref{intphas}) is evaluated with a
standard Monte-Carlo technique which entails following classical
trajectories in phase space.

The electrons are described immediately after absorption by the
distribution
\begin{equation}
\label{phase}
\rho(\Gamma) = \mathscr{N} \delta(\vec{r}_1) \rho_2(\vec{r}_2,\vec{p}_2)
\end{equation}
where $\mathscr{N}$ is a normalization constant.  The primary electron
absorbs the photon which has an energy $\hbar\omega$.  With
$\delta(\vec{r}_1)$ we demand the absorption to occur at the nucleus,
an approximation which becomes exact in the limit of high photon
energy \cite{kabir}.  This approximation significantly reduces the
initial phase space volume to be sampled.  Regularized coordinates
\cite{ks1,ks2} are used to avoid problems with electron trajectories
starting at the nucleus ($\vec{r}_1=0$).

The function $\rho_2 (\vec r_2, \vec p_2)$ describing the secondary
electron in Eq.~(\ref{phase}) is given by
\begin{equation}
\rho_2 (\vec r_2, \vec p_2)=\mathscr{W}_{\psi}
(\vec{r}_2,\vec{p}_2) \delta(\varepsilon^{\rm in}_2-\varepsilon_B).
\end{equation}
It is obtained by calculating the
Wigner distribution, $\mathscr{W}_{\psi} (\vec{r}_2,\vec{p}_2)$, of
the orbital $\psi(\vec{r}_2) = \Psi_0(\vec{r}_1=0,\vec{r}_2)$ for a
choice of initial wavefunction $\Psi_0$, and restricting the initial
energy of the secondary electron, $\varepsilon^{\rm in}_2$, to an
energy shell $\varepsilon_B$.  In the KO mechanism the initial state
correlation is not important so we take the independent particle
wavefunction $\Psi_0 (\vec{r}_1,\vec{r}_2) = (Z^3_{\rm eff} / \pi )
\exp(-Z_{\rm eff}(\vec r_{1}+\vec r_{2}))$ with effective charge
$Z_{\rm eff} = Z-5/16$.  From this choice follows
$\varepsilon_B = -Z_{\rm eff}^2/2$ and $\varepsilon^{\rm in}_2 =
p_2^2 / 2 - Z_{\rm eff} / r_2$.

\begin{figure}
\psfrag{x axis}{excess energy (eV)}
\psfrag{y axis}{$\sigma^{++}/\sigma^+$ (\%)}
\includegraphics[width=.85 \columnwidth]{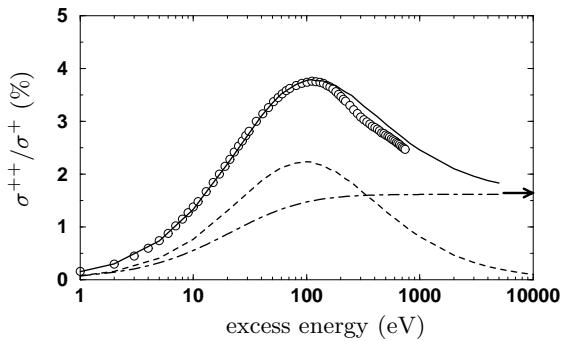}
\caption{\label{fig:rat}Photoionization double-to-single ratio.
Circles: benchmark experimental data (Samson {\it et al.}
\cite{samson98}).  Full line: complete theoretical result.  Dashed
line: knock-out mechanism only.  Chained line: shake-off mechanism
only.  The arrow indicates the asymptotic ratio ($\sim 1.645$\%).}
\end{figure}

The double-to-single ratio in the absence of the SO mechanism is simply
given by
\begin{equation}
\label{ratko}
R_{\text{KO}}=P_{\text{KO}}^{++}/(1-P_{\text{KO}}^{++}).
\end{equation}
In Fig.~\ref{fig:rat} we show $R_{\text{KO}}$ as a function of the
excess energy $E$ (dashed line).  The shape is characteristic of an
impact ionization process \cite{ff}.  For high energies the primary
electron moves away so quickly that there is no time to transfer
energy to the secondary electron, $R_{\text{KO}}$ thus drops to zero
as expected.  The non-zero asymptotic ratio (indicated by an arrow in
Fig.  \ref{fig:rat}) is due to SO which we describe next.
\begin{figure}
\psfrag{x axis}{$\varepsilon$~(eV)}
\psfrag{y axis}{${\rm d}P^{++}/{\rm d}\varepsilon$~($10^{-3}$\,eV)}
\includegraphics[width=\columnwidth]{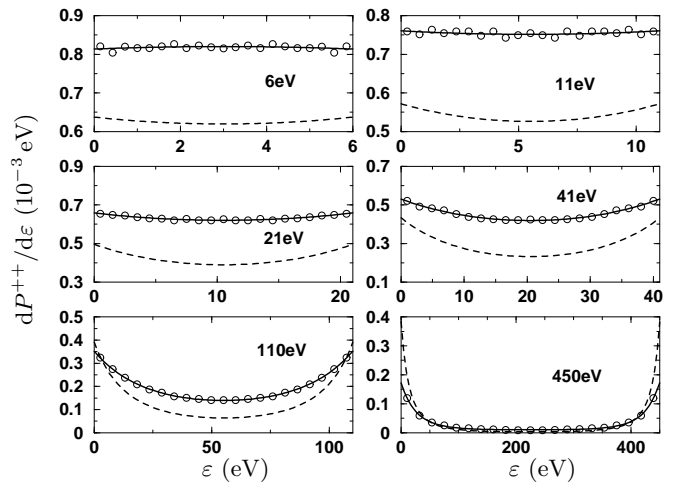}
\caption{\label{fig:diff1}Differential probabilities for separate
knock-out and shake-off mechanisms for a number of excess energies.
Circles: knock-out mechanism results from binning. Solid lines: fits
through circles. Dashed lines: shake-off mechanism results. See text for
details.}
\end{figure}

In contrast to KO the shake mechanism is inherently non-classical in
nature.  Moreover, initial state correlations are important for
shake-off.  As a generalization of the standard formula for SO
\cite{dalgarno}, \AA berg gave an expression for the probability to
find the shake electron in state $\phi_{\alpha}$
at any excess energy \cite{aberg},
\begin{equation}
\label{eqsh1}
P^{\nu}_{\alpha}=
{\left | \langle \phi_{\alpha}|
\psi^{\nu} \rangle \right |^2 }/
{\langle {\psi}^{\nu}|\psi^{\nu}\rangle},
\end{equation}
with
\begin{equation}
\label{eq2}
{\psi}^{\nu}(\vec{r}_{2})=\int {\rm d}^3 r_{1} \nu^\ast (\vec{r}_1)
\Psi_0 (\vec{r}_1,\vec{r}_{2}),
\end{equation}
where $\nu(\vec{r}_{1})$ is the wavefunction of the primary electron
after
it has left the atom.  If it was in an $s$-state before the absorption
it is in a $p$-state afterwards.  The secondary (shake) electron does
not change its angular momentum.  It can be found with probability
$P_{\alpha}$ in an hydrogenic eigenstate of the bare nucleus, being 
either bound
($\alpha=n_2$), or in the continuum ($\alpha=\varepsilon$).  As for KO
we assume that the primary electron absorbs the photon at the nucleus.
In this situation we do not need to know $\nu(\vec r_{1})$ but can
simply replace
${\psi}^{\nu}(\vec{r}_2)$ by $\Psi_0 (\vec{r}_1=0,\vec{r}_2)$ in
Eq.~(\ref{eq2}).  We may further simplify the calculation of shake-off
for practical applications in two-electron atoms by taking for
$\Psi_0 (\vec{r}_1=0,\vec{r}_2)$ a normalized hydrogenic wavefunction
$\phi_{1s}^{Z_{\rm SO}}(\vec{r}_2)$ where the correlations have been
`absorbed' into an effective shake charge $Z_{\rm SO}$ \cite{suric}.
For $Z_{\rm SO} \approx 2-0.51$ the
exact asymptotic ratio $R_{\infty} = 0.01645$ \cite{forrey, krivec} is
reproduced. We have found little
difference for the shake probability as a function of excess energy
between this simple ansatz and a fully correlated Hylleraas
wavefunction \cite{stewart} for $\Psi_{0}$.

The shake-off probability of \eq{eqsh1} reduces now to
\begin{equation}
\label{eq4}
P_{\alpha} =\left|\langle\phi_{\alpha} |\phi^{Z_{\rm
SO}}\rangle\right|^2\,.
\end{equation}
The total
double ionization probability from shake-off
at finite energies $E$ is given by integrating expression (\ref{eq4}) 
over the energy
$\varepsilon$ of the shake
electron in the continuum ($\alpha\equiv \varepsilon $),
\begin{eqnarray}
\label{eq5}
P_{\rm SO}^{++}(E) &=& \int^E_0 {\rm d}\varepsilon \, P_{\varepsilon}.
\end{eqnarray}

\begin{figure}
\psfrag{x axis}{$\varepsilon/E$}
\psfrag{y axis}{${\rm d}\sigma^{++}/{\rm d}\varepsilon$ normalized to 1 
for $\varepsilon=0$}
\includegraphics[width=.9 \columnwidth]{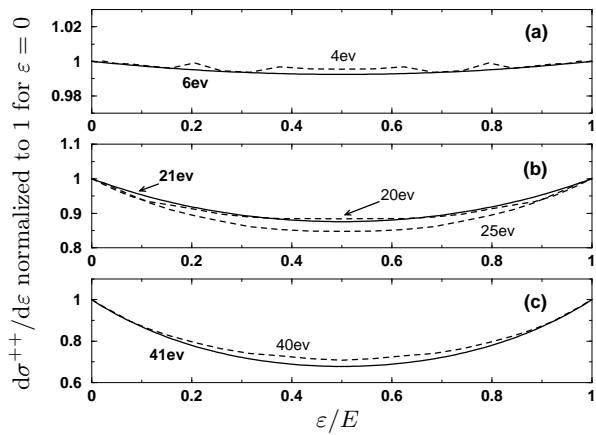}
\caption{\label{fig:diff2}Singly differential cross sections
normalized to 1 for $\varepsilon=0$.  Solid lines: our complete
theoretical results at excess energies of (a) 6~eV, (b) 21~eV, (c)
41~eV. Dashed lines: new results of Colgan {\it et al.}
\cite{colgan01} at excess energies of (a) 4~eV, (b) 20~eV and 25~eV,
(c) 40~eV.}
\end{figure}

The photoionization ratio when {\it only} the SO mechanism is taken
into account (same as Eq.~(\ref{ratko}) but for SO) is shown in Fig.
\ref{fig:rat} (chained line).  The ratio rises slower than the KO
mechanism result up to an energy of around $100$~eV where
the KO ratio reaches its maximum value.  The SO ratio continues to
rise until at a couple of hundred eV it moves more slowly up toward
the asymptotic value.  An interesting feature of the plot is where the 
KO and SO results cross at an excess energy of $\sim 350$~eV.

To obtain more insight into the two mechanisms we calculate the
differential probabilities ${\rm d} P^{++}_{\rm X}/{\rm d}
\varepsilon$, where ${\rm X}$ stands for either SO or KO. In our
classical model of the KO mechanism we divide the interval of values
for $\varepsilon$ which corresponds to double escape ($0\le
\varepsilon \le E$) into $N$ equally sized bins (we take $N=21$) and
work out the differential probability by finding the trajectories
which fall into the bins.  For the SO mechanism the probability per
energy unit, $P_{\varepsilon}$ in Eq.~(\ref{eq5}),
already gives the differential probability. Since the electrons are
indistinguishable the differential
probabilities must be symmetrized about the equal energy sharing point
($\varepsilon  = E- \varepsilon = E/2$),
\begin{equation}
\left.\frac{{\rm d} P_{\rm X}^{++}}{{\rm d}\varepsilon}\right|_{\rm sym}=
\frac{1}{2}\left( \frac{{\rm d} P^{++}_{\rm X}(\varepsilon, E)}{{\rm d}
\varepsilon} +
\frac{{\rm d} P^{++}_{\rm X}(E-\varepsilon,E)}{{\rm d} \varepsilon}
\right).
\end{equation}

\begin{figure}[t]
\psfrag{x axis}{$\varepsilon$~(eV)}
\psfrag{y axis}{${\rm d}\sigma^{++}/{\rm d}\varepsilon$~(kb/eV)}
\includegraphics[width=.9 \columnwidth]{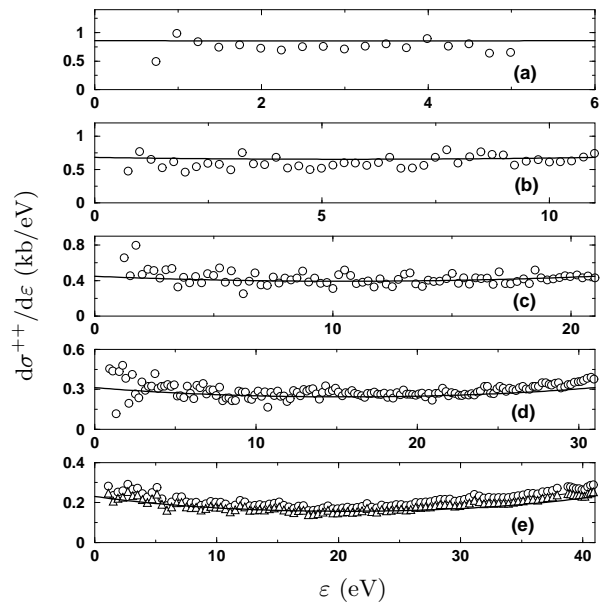}
\caption{\label{fig:diff3}Absolute singly differential cross sections.
Solid lines: our theoretical results at excess energies of (a) 6~eV,
(b) 11~eV, (c) 21~eV, (d) 31~eV, (e) 41~eV. Circles: recalibrated (see
text) experimental data of Wehlitz {\it et al.} \cite{wehlitz91} at
the same excess energies apart from (a) which is at 5~eV. The
triangles in (e) additionally show the Wehlitz data renormalized to
the $\sigma^{++}(41~{\rm eV})$ of Samson \cite{samson98}.}
\end{figure}
In the case of low excess energy (6~eV) we find a slightly concave shape
for the KO distribution, see Fig.~\ref{fig:diff1}. This implies a
preference for  equal energy sharing,
the typical behavior close to threshold \cite{rost}.
The SO result in contrast displays a slightly convex shape which
becomes flat as $E \rightarrow +0$.  Unequal energy sharing is always
preferred by SO since the photoelectron is fast with respect to
the secondary electron.  For all higher excess energies shown both
mechanisms display a convex form.

SO may be viewed as an additional quantum contribution to the 
quasi-classically calculated double
photoionization given by KO. This means that the full result is given by
\begin{equation}
\label{jointdiff}
\frac{{\rm d} \sigma^{++}}{{\rm d} \varepsilon} = \sigma_{\rm
abs}\left(\frac{{\rm d}
P^{++}_{\rm KO}}{{\rm d} \varepsilon} + \frac{{\rm d}
P^{++}_{\rm SO}}{{\rm d} \varepsilon}\right).
\end{equation}
Integration over $\varepsilon$ yields the total double ionization cross
section,
\begin{equation}
\label{jointint}
\sigma^{++}=\sigma_{\rm abs}(P^{++}_{\rm KO}+P^{++}_{\rm
SO})\equiv\sigma^{++}_{\rm KO}+\sigma^{++}_{\rm SO}.
\end{equation}
    The single ionization cross
section is $\sigma^+ = \sigma_{\rm abs} - \sigma^{++}$ and
   the double-to-single ratio is given by
$R=\sigma^{++}/\sigma^+ =P^{++}/(1-P^{++})$, where $P^{++}=P^{++}_{\rm
KO}+P^{++}_{\rm SO}$.

In Fig.~\ref{fig:rat} we compare the ratio $R$ (solid line) to the
experimental data of Samson \textit{et al.} \cite{samson98}.  For
excess energies up to $200$~eV we find an excellent agreement.  In the
energy regime where the two contributions are of the same size there
is a deviation between experiment and our result (at worst 8\%).
Exactly in this situation any interference which exists between SO and KO
would show its largest effect. The deviation we find may be due to such 
an
interference
which we cannot account for since we determine KO quasi-classically.
At higher energies the difference decreases again (already visible in
the plot) until at very high energies our result reproduces the
asymptotic ratio.

   Knowing that the differential probabilities for KO and SO enter the
   full ionization probability with the same weight we can assess the
   relative importance of both contributions at different energies.
   From  Fig.~\ref{fig:diff1} one sees that at 110~eV SO has
   become more important than KO for highly unequal
   energy sharings.  As energy is increased to 450~eV SO begins to 
dominate regions of
   unequal energy sharing. On the other hand,  KO is higher at
   equal energy sharing  for all excess energies $E$.

Fig.~\ref{fig:diff2} shows that our singly differential cross sections
(SDCS) agree well with the recent {\it ab initio} theoretical
results of Colgan {\it et al.} \cite{colgan01}.  We note that the
results of Proulx and Shakeshaft \cite{shake} show a concave shape for
excess energies below 20~eV. This is in disagreement with our results
which are convex down to 6~eV. In Fig.~\ref{fig:diff3} we compare our
absolute SDCS to the experimental data of Wehlitz {\it et al.}
\cite{wehlitz91} which has been recalibrated using the values of the
photoabsorption cross section of Samson {\it et al.} \cite{sam94}.  In
their original work Wehlitz {\it et al.} normalized their SDCS using a
photoabsorption cross section \cite{mars} which is now known to
overestimate the 5-41~eV range by 9 to 16\%.  In addition, the
photoionization ratio they measured at 41~eV is $\sim 17$\% higher
than the Samson {\it et al.} \cite{samson98} data and so we
renormalize Fig.~\ref{fig:diff3}(e) to take this into account.

\begin{figure}
\psfrag{x axis}{scaled energy}
\psfrag{y axis}{$P^{++}_{\rm KO}$}
\includegraphics[width=.85 \columnwidth]{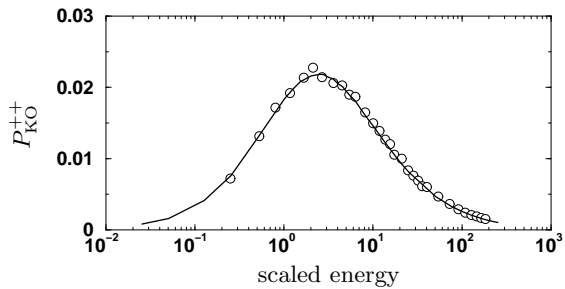}
\caption{\label{fig:impactHe}$P^{2+}_{\rm KO}$ (solid line) as a
function of the scaled energy $E/E_{\rm B}'$ compared to the cross
section for electron impact ionization of He$^+$ \cite{He+exp} (circles)
as a function of $E/E_{\rm B}$ (see text).
Additionally the impact ionization data has been multiplied by a factor
$C=4.67\times 10^{15}~$cm$^{-2}$ to make the maxima of both curves the
same height. ($1/C$ may be interpreted as the geometric cross section.)}
\end{figure}

Our approach not only facilitates the
   calculation of double photoionization, it also offers considerable
   insight into the physical process, e.g., concerning the similarity
   with electron impact ionization of He$^+$ \cite{samsonxx}. Indeed, we
can show
   that impact ionization may be viewed as the KO part of double 
photoionization (Fig.~\ref{fig:impactHe}). The only difference is that
   impact ionization sees a He$^+$ hydrogenic target electron with
   binding energy $E_{\rm B}= -Z^2/2, Z = 2$, while the KO process 
involves
   a bound electron with energy $E_{\rm B}' = -Z_{\rm eff}^2/2$. One
   may thus say that both processes differ only slightly, namely in the
energy scale set
   by the respective bound electron.

   We conclude that the separate formulation and calculation
   of knock-out and shake-off offers an
   accurate description of double photoionization. In principle this
   approach can be extended to describe angular differential cross
   sections. We have used a description in terms of the simplest
   wavefunctions possible having in mind to tackle three electron
   problems in a similar way.

It is a pleasure to thank R. Wehlitz and J. Colgan for providing us with
their results.
T.S. thanks Andreas Becker and Thomas Pattard for valuable discussions.
Financial support by the DFG within the Gerhard Hess-program is
gratefully acknowledged.

\end{document}